\documentclass[
  preprintnumbers
]{revtex4}

\usepackage{amsthm}
\usepackage{amsmath}
\usepackage{amssymb}
\usepackage{bbold}
\usepackage{graphicx}
\usepackage{subfigure}

\usepackage{enumerate}

\usepackage{tikz}
\usepgflibrary{shapes.geometric}
\usetikzlibrary{arrows,decorations.pathreplacing}

\newtheorem{theorem}{Theorem}
\newtheorem{lemma}{Lemma}

\theoremstyle{definition}
\newtheorem{definition}{Definition}

\newcommand{\ket}[1]{|#1\rangle}
\newcommand{\bra}[1]{\langle #1 |}

\newcommand{\braket}[2]{\langle #1 |#2 \rangle}

\newcommand{\proj}[1]{\ket{#1}\bra{#1}}

\newcommand{\ot}[0]{\otimes}

\newcommand{\beq}{\begin{equation}}
\newcommand{\eeq}{\end{equation}}
\newcommand{\best}{\begin{equation*}}
\newcommand{\eest}{\end{equation*}}

\newcommand{\idmap}{{\rm id}}

\newcommand{\supp}{\textrm{supp}}

\DeclareMathOperator{\Tr}{Tr}

\def\I{\mathbb{1}}

\begin{document}

\title{Local Broadcasting of Quantum Correlations}

\author{Marco Piani}
\affiliation{SUPA and Department of Physics, University of Strathclyde, Glasgow G4 0NG, UK}

\begin{abstract}
Operations that are trivial in the classical world, like accessing information without introducing any change or disturbance, or like  copying information, become non-trivial in the quantum world. In this note we discuss several limitations in the local redistributing  correlations, when it comes to dealing with bipartite quantum states. In particular, we focus on the task of local broadcasting, by discussing relevant no-go theorems, and by quantifying the non-classicality of correlations in terms of the degree to which local broadcasting is possible in an approximate fashion.
\end{abstract}

\maketitle

\section{Cloning, broadcasting, and local broadcasting}

\subsection{Cloning}

A key aspect of quantum information, which strongly differentiates it from classical information, is the inability to freely copy quantum states. Suppose we are given a quantum system $S_1$ in an unknown state $\ket{\psi}$, and that we want to put -- equivalently, prepare -- another system $S_2$ (let us say, of similar physical nature, and initially in some fiducial state $\ket{0}$) in the same state, without changing the state of the system that was given to us. Thinking in terms of a unitary evolution $U$, what we want to accomplish is the following:
\beq
\label{eq:unitarycloning}
U\ket{\psi}_{S_1}\ket{0}_{S_2}\ket{0}_{E} = \ket{\psi}_{S_1}\ket{\psi}_{S_2}\ket{\xi_\psi}_E,
\eeq
where $E$ is an ancillary system that we may want to use in the process, initially prepared in some fiducial state $\ket{0}_E$ independent of $\ket{\psi}$ and ending up in a state $\ket{\xi_\psi}$, which potentially depends on $\ket{\psi}$. Consider now two known  states $\ket{\psi}$ and $\ket{\psi'}$, for both of which we assume \eqref{eq:unitarycloning} to hold for the same unitary $U$. 
By taking the inner product of the left-hand and of the right-hand sides of the two occurrencies (one for $\psi$, and one for $\psi'$) of \eqref{eq:unitarycloning}, and using the fact that $U$ preserves inner products, we arrive at the relation 
\beq
\braket{\psi'}{\psi} = \braket{\psi'}{\psi}^2\braket{\xi_{\psi'}}{\xi_\psi}.
\eeq
Given that the modulus of the inner product of two normalized vector states is always less or equal to 1, the latter relation can be satisfied only if $|\braket{\psi'}{\psi}|$ is either 0 ($\psi$ and $\psi'$ are orthogonal) or 1 ($\psi$ and $\psi'$ are the same). This is the content of the no-cloning theorem~\cite{wootters1982single,DIEKS1982271}, which says that, within the quantum formalism, there is no physical process able to clone pure quantum states that are not orthogonal.  

In the general case where one adopts the formalism of density matrices and channels, the cloning of a state $\rho$ of a system $S$ by means of an $S\rightarrow S_1S_2$ channel $\Lambda$ corresponds to the request
\beq
\Lambda[\rho_S] = \rho_{S_1}\otimes\rho_{S_2},
\eeq
where the introduction of a system onto which to copy the state and the possibility of using an ancillary system in the process are already taken into account by the quantum channel formalism. It is useful to recall that every quantum channel $\Lambda_{S\rightarrow S'}$ from a system $S$ to a system $S'$ (of potentially different dimensionality) can be seen as the result of an isometry $V_{S\rightarrow S'E}$ from $S$ to a combined system $S'E$, followed by the tracing out of $E$:
\beq
\label{eq:stinespring}
\Lambda_{S\rightarrow S'}[\rho_S] = \Tr_{E}\left(V_{S\rightarrow S'E} \rho_S V_{S\rightarrow S'E}^\dagger\right)
\eeq
This is known as Stinespring or isometric dilation of quantum channels~\cite{wilde2013quantum}. In the case of cloning, the output system $S'$ consists of two copies of the input system $S$, that is,  $S'=S_1S_2$.

Consider the fidelity between two states $\rho$ and $\sigma$~\cite{nielsen2010quantum},
\beq
F(\rho,\sigma):=\Tr\left(\sqrt{\sqrt{\rho}\sigma\sqrt{\rho}}\right)=\|\sqrt{\rho}\sqrt{\sigma}\|_1.
\eeq
In the rightmost expression, $\|\cdot\|_1$ indicates the 1-norm (also called trace norm), $\|X\|_1:=\Tr\sqrt{XX^\dagger}=\Tr\sqrt{X^\dagger X}$. Such an expression shows explicitly that the fidelity is symmetric in $\rho$ and $\sigma$. The fidelity satisfies $0\leq F(\rho,\sigma)\leq1$, with $F(\rho,\sigma)=0$ if and only if $\rho$ and $\sigma$ are orthogonal, and $F(\rho,\sigma)=1$ if and only if $\rho=\sigma$. Furthermore, it is multiplicative on tensor states,
 \beq
 F(\rho\otimes\rho',\sigma\otimes\sigma')=F(\rho,\sigma)F(\rho',\sigma'),
 \eeq
 and it is monotone under quantum channels, i.e.,
 \[
 F(\Gamma[\rho],\Gamma[\sigma])\geq F(\rho,\sigma),
 \]
 for any pair of states $\rho,\sigma$ and any quantum channel $\Gamma$.
Suppose now that $\rho$ and $\rho'$ can each be cloned by the action of the same $\Lambda$.
Then we have
\beq
\label{eq:fidelitycloning}
F(\rho,\rho') \leq F(\Lambda[\rho] ,\Lambda[\rho'] )= F(\rho\ot\rho ,\rho'\ot\rho' )=F(\rho,\rho')F(\rho,\rho').
\eeq
The inequality is due to the monotonicity of the fidelity under quantum channels; the first equality is just the hypothesis that $\Lambda$ is able to clone both $\rho$ and $\rho'$; the second equality is due to the multiplicativity of fidelity on tensor states. Since $0\leq F(\rho,\rho')\leq 1$, Eq. \eqref{eq:fidelitycloning} can hold only in two situations: either $F(\rho,\rho')=0$ (the states are orthogonal) or $F(\rho,\rho')=1$ (the two states are actually the same state).  Notice that in both  latter circumstances the two states commute, $[\rho,\rho']=0$, and they do so in a trivial way. Thus, we find that, when it comes to cloning, the conditions for it to be possible do not vary when moving from pure states to the consideration of general mixed states.


\subsection{Broadcasting}

In the general framework of density operators, though, one can relax the condition of cloning to that of `broadcasting', for which we only require that $\tilde{\rho}_{S_1S_2}=\Lambda_{S\rightarrow S_1S_2}[\rho_{S}]$ is such that its marginals $\tilde{\rho}_{S_1}=\Tr_{S_2}(\tilde{\rho}_{S_1S_2})$ and $\tilde{\rho}_{S_2}=\Tr_{S_1}(\tilde{\rho}_{S_1S_2})$ satisfy
\beq
\label{eq:broadcastingcond}
\tilde{\rho}_{S_1} = \tilde{\rho}_{S_2} = \rho_{S}.
\eeq
This means that we require that `having two copies of $\rho$' is only achieved at the level of reduced states of the output systems, and the copies are not necessarily independent, that is, we allow $\tilde{\rho}_{S_1S_2}\neq\tilde{\rho}_{S_1}\otimes\tilde{\rho}_{S_1}$.
It should be clear that mixed density matrices that are the convex combinations of a fixed set of orthonormal pure states, can be broadcast in the above sense. Indeed, for any fixed set $\{\ket{\psi_i}\}$ of states respecting $|\braket{\psi_i}{\psi_j}|=\delta_{ij}$, there is a channel $\Lambda$ such that
\beq
\label{eq:singlebroadcast}
\Lambda\left[\sum_i p_i \proj{\psi_i}_S\right] = \sum_i p_i \proj{\psi_i}_{S_1} \otimes \proj{\psi_i}_{S_2},
\eeq
for any
\emph{any} probability distribution $\{p_i\}$. Indeed, it is enough to consider a channel $\Lambda$ that clones the orthogonal pure states $\{\ket{\psi_i}\}$---something we well know to be possible---and to exploit the linearity of the action of a channel. One immediately checks that the state on the right-hand side of \eqref{eq:singlebroadcast} is such that the condition \eqref{eq:broadcastingcond} is satisfied with respect to $\rho_S=\sum_i p_i \proj{\psi_i}_S$. What is less trivial is that this is the only case where broadcasting of mixed quantum states is possible; this fact is captured by the no-broadcasting theorem~\cite{barnum1996noncommuting}.
\begin{theorem}
\label{thm:nobroad}
Two mixed states $\rho$ and $\rho'$ can be broadcast simultaneously if and only if they admit a spectral decomposition with the same eigenvectors, that is, if and only if they commute, $[\rho,\rho'] = 0$.
\end{theorem}
Notice that, this theorem is immediately extended to a collection of states, since pairwise commutation of Hermitian operators implies joint commutation.

\subsection{Local broadcasting}

It is worth stressing that, for a fixed and known state $\rho$, there is no problem with `cloning', as knowing the state allows one to create as many copies of it as one wants. Nonetheless, limitations kick in again even for a single state if we consider restrictions on the operations, like allowing only local operations in a distributed setting. Then, the broadcasting of a single state of a distributed system cannot be performed arbitrarily. For example, consider the case where Alice and Bob would like to clone the maximally entangled state $\ket{\psi^+}_{AB} =\frac{1}{\sqrt{2}}(\ket{00}_{AB}+\ket{11}_{AB})$. Notice that, since we are considering a pure state, we are dealing with actual cloning, not with broadcasting, that is, the target state is composed of independent copies. If Alice and Bob could use global operations, with no limit on what they can do across their laboratories, then Alice  and Bob could certainly produce two copies of $\ket{\psi^+}_{AB}$---even from scratch, without using the fact that they shared one copy of the state to begin. If instead they can only implement fully local quantum channels of the form $\Lambda_{AB} = \Lambda_A \otimes \Lambda_B$, then they cannot transform one copy of $\ket{\psi^+}$ into two copies of it, because they cannot increase the entanglement they share. Actually, the task would be impossible even if they were allowed to communicate classically, that is, allowed to apply Local Operations aided by Classical Communication (LOCC)~\cite{revent}.

Is there a general no-go theorem for local broadcasting, that goes beyond considerations related to entanglement? Yes, there is! In Ref.~\cite{piani2008no} the following was proved.
\begin{theorem}
\label{thm:nolocal}
Let $\rho_{AB}$ be a bipartite state. There exist local maps $\Lambda_{A\rightarrow A_1A_2}$ and $\Gamma_{B\rightarrow B_1B_2}$ such that
\beq
\tilde{\rho}_{A_1A_2B_2B_2}=(\Lambda_{A\rightarrow A_1A_2}\otimes\Gamma_{B\rightarrow B_1B_2})[\rho_{AB}]
\eeq
satisfies
\beq
\label{eq:localbroadcond}
\tilde{\rho}_{A_1B_1}=\tilde{\rho}_{A_2B_2}=\rho_{AB},
\eeq
if and only if $\rho_{AB}$ is classical-classical, that is,
\beq
\rho_{AB} = \sum_{ij} p_{ij} \proj{a_i}_{A}\ot\proj{b_j}_{B},
\eeq
with $\{\ket{a_i}_A\}$ ($\{\ket{b_j}_{B}\}$) some orthonormal basis for $A$ ($B$).
\end{theorem}

The previous result can be considered a no-local-broadcasting theorem, more precisely a two-sided no-local broadcasting theorem, that spells out a limitation about broadcasting in a distributed setting, when only local operations are allowed. In the next sections we will prove it, and even prove  a one-sided version of it. Before we do that, we will recall some entropic quantifiers of distinguishability of quantum states and of correlations, and their relation with quantum recoverability.

\section{Entropy, relative entropy, mutual information, and conditional mutual information}

\subsection{Entropy}
We begin by recalling the notion of von Neumann entropy~\cite{nielsen2010quantum}.
\begin{definition}
The von Neumann Entropy a quantum state $\rho$ is a quantifier of how mixed $\rho$ is, and is defined as
\beq
S(\rho):=-\Tr(\rho\log_2\rho).
\eeq
\end{definition}

In the following we will sometimes consider the entropy of subsystems. The entropy $S(X)_\rho$ of a subsystem $X$ of a bipartite system $XY$ in a global state $\rho=\rho_{XY}$ is defined as
\beq
S(X)_\rho := S(\rho_X),
\eeq
where $\rho_X$ is the reduced state of $X$. Notice that, when considering a multipartite system, we can always think of the bipartition into, on one side,  the system of interest and, on the other side, all the other systems.

\subsection{Relative entropy}
Relative entropy is a quantifier of the distinguishability of two quantum states~\cite{nielsen2010quantum}.
\begin{definition}
The relative entropy  $S(\rho\|\sigma)$  between a quantum state $\rho$ and a quantum state $\sigma$ is defined as
\beq
\label{eq:relativentropy}
S(\rho\|\sigma) =
\begin{cases}
\Tr(\rho \log_2\rho)-\Tr(\rho\log_2\sigma) & \supp(\rho)\subseteq\supp(\sigma) \\
+\infty & \textrm{otherwise}.
\end{cases}
\eeq
Notice that the relative entropy is not symmetric in the two arguments $\rho$ and $\sigma$. In the following we will always consider the first case, $\supp(\rho)\subseteq\supp(\sigma)$.
\end{definition}

It holds that $S(\rho\|\sigma)\geq0$, with equality, $S(\rho\|\sigma)=0$, if and only if $\rho=\sigma$. Furthermore the relative entropy is monotone under channels, that is,
\beq
\label{eq:DPIrelative}
S(\rho\|\sigma)\geq S(\Gamma[\rho]\|\Gamma[\sigma]),
\eeq
for any pair of states $\rho,\sigma$ and any quantum channel $\Gamma$.
The latter relation is often called  the `data-processing inequality' for relative entropy; operationally, it follows directly form the interpretation of relative entropy as measure of distinguishability between the two states~\cite{vedralrelative}.

\subsection{Mutual information}
Mutual information is a quantifier of correlations encoded in a bipartite quantum state, that can be understood as the relative entropy between the state and the product of its marginals~\cite{nielsen2010quantum,vedralrelative}.
\begin{definition}
The mutual information $I(A:B)_\rho$ between systems $A$ and $B$ in a state $\rho=\rho_{AB}$ can be defined as
\beq
\label{eq:mutualinfo}
I(A:B)_\rho:=S(\rho_{AB}\|\rho_A\ot\rho_B)=S(A)_\rho+S(B)_\rho-S(AB)_\rho.
\eeq
\end{definition}
The rightmost side of \eqref{eq:mutualinfo} comes from computing the relative entropy for the specific choice of states $\rho$ and $\sigma$ in \eqref{eq:relativentropy}, and proves that mutual information is symmetric under the exchange of $A$ and $B$. A consequence of the data-processing inequality for relative entropy is that mutual information is monotone under fully local quantum channels of the form $\Lambda_{A\rightarrow A'} \otimes \Lambda_{B\rightarrow B'}$: if $\rho'=\rho'_{A'B'}=(\Lambda_{A\rightarrow A'} \otimes \Lambda_{B\rightarrow B'})[\rho_{AB}]$, then $I(A:B)_{\rho'} \leq I(A:B)_\rho$. Notice that, since $\Lambda_{A\rightarrow A'} \otimes \Lambda_{B\rightarrow B'} = (\Lambda_{A\rightarrow A'} \otimes \idmap_B) \circ (\idmap_A \otimes \Lambda_{B\rightarrow B'})$, monotonicity of mutual information under fully local quantum channel is equivalent to the monotonicity of mutual information under both channels that act non-trivially only on $A$, i.e., of the form $\Lambda_{A\rightarrow A'}  \otimes \idmap_B$, and channels that act non-trivially only on $B$.

\subsection{Conditional mutual information}
Let us consider a tripartite system $ABC$ in a state $\rho=\rho_{ABC}$.
It is useful to introduce the notion of conditional mutual information between $A$ and $C$, conditioned on $B$, defined as~\cite{nielsen2010quantum,wilde2013quantum}
\beq
\label{eq:condmutual}
\begin{aligned}
I(A:C|B)_\rho
&:= I(A:BC)_\rho -I(A:B)_\rho \\
&= S(AB)_\rho+S(BC)_\rho-S(ABC)_\rho-S(B)_\rho.
\end{aligned}
\eeq
The last expression for the conditional mutual information in~\eqref{eq:condmutual} proves that $I(A:C|B)_\rho$ is symmetric between $A$ and $C$; indeed, it holds
\[
I(A:C|B)_\rho =  I(A:BC)_\rho -I(A:B)_\rho =  I(AB:C)_\rho -I(C:B)_\rho. 
\]
Conditional mutual information $I(A:C|B)_\rho$ is non-negative, a fact known also as the strong subadditivity of von Neumann entropy~\cite{nielsen2010quantum,wilde2013quantum}, and of the foremost importance in quantum information theory. It has an interpretation as the amount of correlations, as measured by mutual information, lost between $A$ and $BC$ when $C$ gets discarded.

Notice that, if we consider the tripartite state $\rho_{AB'E}=V_{B\rightarrow B'E} \rho_{AB} V_{B\rightarrow B'E}^\dagger$, that arises from the isometry step in the implementation of an arbitrary local channel $\Lambda_{B\rightarrow B'}$, the mutual information between $A$ and $B'E$ is the same as the mutual information between $A$ and $B$, before the action of the channel. On the other hand, the mutual information after the action of the channel, is the mutual information between $A$ and $B'$, after having discarded $E$. Thus we see that the fact that conditional mutual information 
is non-negative is equivalent to the fact that mutual information is monotone under fully local operations. 

\section{Quantum recoverability}
\label{sec:quantumrecoverability}

The data processing inequality \eqref{eq:DPIrelative} can be refined, and linked to the issue of the recoverability of the action of the quantum channel $\Gamma$ on $\rho$~\cite{wilde2015recoverability,junge2015universal}.
\begin{theorem}
\label{thm:fawzirenner}
Given two states $\rho$ and $\sigma$, and a channel $\Gamma$, there is a recovery channel $R=R_{\sigma,\Gamma}$ that depends only on $\sigma$ and $\Gamma$ such that
\begin{gather}
\label{eq:fawzirenner} S(\rho||\sigma)- S(\Gamma[\rho]\|\Gamma[\sigma])\geq - \log_2 F^2(\rho,(R\circ \Gamma) [\rho]) \\
(R\circ \Gamma) [\sigma] = \sigma.
\end{gather}
\end{theorem}

Notice that the right-hand side of \eqref{eq:fawzirenner} is non-negative, so that indeed \eqref{eq:fawzirenner} constitutes a strengthening of \eqref{eq:DPIrelative}. In Theorem~\ref{thm:fawzirenner}, the recovery channel $R$ always recovers $\sigma$ from $\Gamma[\sigma]$ perfectly. On the other hand, how well $R$ recovers $\rho$ from $\Gamma[\rho]$---that is, how large the fidelity  $F(\rho,(R\circ \Gamma) [\rho])$ can be---depends on the decrease of the relative entropy under the action of $\Gamma$: the fidelity is large---close to $1$---if the decrease in the relative entropy is small, since \eqref{eq:fawzirenner} is equivalent to
\[
 F(\rho,(R\circ \Gamma) [\rho])  \geq 2^{-\frac{1}{2}\left(S(\rho||\sigma)- S(\Gamma[\rho]\|\Gamma[\sigma])\right)}.
\]

Notice that the state $\rho$ is perfectly recovered by $R$ from $\Gamma [\rho]$---that is, the fidelity of $(R\circ \Gamma) [\rho]$ with $\rho$ is equal to $1$---if there is no decrease of the relative entropy. The case of equality in the left-hand side of \eqref{eq:fawzirenner} had already been considered by Petz, who was able to provide an explicit form for a perfect recovery map $R^P=R^P_{\sigma,\Gamma}$~\cite{petz1986sufficient,petz1988sufficiency,hayden2004structure} for such a case:
\beq
\label{eq:petzmap}
R^P_{\sigma,\Gamma} [\tau] = \sigma^{1/2}\Gamma^\dagger\left[(\Gamma[\sigma])^{-1/2}\tau(\Gamma[\sigma])^{-1/2}\right]\sigma^{1/2}.
\eeq
Here $\Gamma^\dagger$ is the map dual to $\Gamma$, i.e., such that $\Tr(X^\dagger \Gamma[Y]) = \Tr((\Gamma^\dagger[X])^\dagger Y)$ for all $X,Y$. In particular, since $\Gamma$ is a channel that admits a Kraus decomposition, $\Gamma[Y] = \sum_i K_i Y K_i^\dagger$, with Kraus operators $K_i$, then $\Gamma^\dagger$ acts as follows: $\Gamma^\dagger[X] = \sum_i K_i^\dagger X K_i$. One verifies that $R^P$ is completely positive and trace-preserving, hence a channel. In the general case covered by Theorem~\ref{thm:fawzirenner}, which considers a non-vanishing decrease in the relative entropy, it has been proven that the recovery channel $R$ can be chosen to have some close connection with the structure of the Petz recovery channel~\cite{wilde2015recoverability,junge2015universal}, but a discussion of the present status in the study of the expression of the best recovery map---in the general case of imperfect recovery---goes beyond the scope of these notes. 

\subsection{Quantum recoverability and mutual information}

Given that mutual information is as a special case of relative entropy, it should be no surprise that Theorem~\ref{thm:fawzirenner} can be specialized to the case of mutual information. Actually, Theorem~\ref{thm:fawzirenner} can be seen as a generalization of a theorem previously derived by Fawzi and Renner about mutual information~\cite{fawzi2014quantum}.
\begin{theorem}
\label{thm:originalfawzirenner}
For any state $\rho=\rho_{ABC}$ there is a recovery channel $R_{B\rightarrow BC}$ such that
\[
F(\rho_{ABC},R_{B\rightarrow BC}[\rho_{AB}])  \geq 2^{-\frac{1}{2}I(A:C|B)_\rho}.
\]
\end{theorem}
Theorem \ref{thm:originalfawzirenner} says that, if correlations between $A$ and $BC$ do not decrease too much because of the loss (that is, the tracing out) of $C$, then the total state $\rho_{ABC}$ can be recovered pretty well by means of a channel $R_{B\rightarrow BC}$ acting on $\rho_{AB}$; more in detail, it is possible to choose such a map so that it only depends on $\rho_{BC}$, and not on the full state $\rho_{ABC}$~\cite{sutter2016universal}.


\section{Proof of the no-local-broadcasting theorem}

We will use what we recalled about quantum recoverability in Section~\ref{sec:quantumrecoverability} to prove the no-local-broadcasting Theorem~\ref{thm:nolocal}; we will do so by leveraging the no-broadcasting Theorem~\ref{thm:nobroad}. We will use an intermediate step that one could call the no-unilocal-broadcasting theorem~\cite{Luo2010,luo2010decomposition}.

\subsection{No-unilocal-brodcasting}

We say that a bipartite state $\rho_{AB}$ can be locally broadcast on $B$ if there exists a local map $\Gamma_{B\rightarrow B_1B_2}$ such that
\beq
\tilde{\rho}_{AB_1B_2}=(\idmap_A\otimes\Gamma_{B\rightarrow B_1B_2})[\rho_{AB}]
\eeq
satisfies
\beq
\tilde{\rho}_{AB_1}=\tilde{\rho}_{AB_2}=\rho_{AB}.
\eeq
The following holds~\cite{Luo2010,luo2010decomposition}.
\begin{theorem}
\label{thm:unilocalbroad}
A bipartite state $\rho_{AB}$ can be locally broadcast on $B$ if and only if $\rho_{AB}$ is classical on $B$, that is,
\beq
\rho_{AB} = \sum_{j} p_{j} \rho^A_j\ot\proj{b_j}_{B}
\eeq
with $\{\ket{b_j}_{B}\}$ some orthonormal basis for $B$.
\end{theorem}
In order to prove Theorem~\ref{thm:unilocalbroad} we will need the following lemma, for which we provide a more direct proof than the one given in Ref.~\cite{luo2010decomposition}.
\begin{lemma}
\label{lem:decomp}
Any bipartite state $\rho_{AB}$ admits a decomposition of the form
\beq
\label{eq:decomp}
\rho_{AB}=\sum_i p_i F^A_i \otimes \rho^B_i
\eeq
with $\{p_i\}$ a probability distribution, $\{F^A_i\}$ a collection of linearly independent operators on $A$, and $\rho^B_i$ normalized states on $B$.
\end{lemma}
Notice that the operators $\{F^A_i\}$ in \eqref{eq:decomp} are in general non-positive; this must be the case, because otherwise the lemma would claim that every bipartite state can be written as the convex combination of tensor products of positive operators, while we know that the latter applies---by definition---only to unentangled states.
\begin{proof} (of Lemma \ref{lem:decomp}) We will consider a minimal informationally complete POVM  on $A$, which means a collection of linearly independent operators $\{E^A_i\}$ that form a valid POVM, i.e., $E^A_i\geq 0$, $\sum_iE^A_i = I_A$, and at the same time constitute a basis for the space of operators on $A$~\cite{definetti2002}. A minimal informationally complete POVM constitutes a quantum frame~\cite{ferrieemerson}. We can then consider a dual frame $\{F_i\}$ to it, that is, a collection of linearly independent operators such that
\[
X_A = \sum_i \Tr(E_i X) F_i\quad \forall X.
\]
Then, one has
\[
\rho_{AB} = \sum_i p_i F_i \otimes \rho^B_i,
\]
with
\beq
\label{eq:condstates}
\rho^B_i = \frac{1}{p_i}\Tr_A(E^A_i\otimes \I_B \rho_{AB})\quad p_i = \Tr_{AB}(E^A_i\ot\I_B\rho_{AB}).
\eeq
Notice that the $\rho^B_i$'s are normalized states, thanks to the fact that $\{E_i\}$ is a POVM; for the same reason, $\{p_i\}$ is a valid probability distribution.
\end{proof}
We are now in the position to prove Theorem~\ref{thm:unilocalbroad}.
\begin{proof} (of Theorem~\ref{thm:unilocalbroad}) That a state classical on $B$ can be locally broadcast on  $B$ is trivial. To prove the other direction, let us consider a map $\Gamma_{B\rightarrow B_1B_2}$ that achieves the local broadcasting of $\rho_{AB}$ on $B$. We are going to prove that  $\Gamma_{B\rightarrow B_1B_2}$ broadcasts the individual states $\rho^B_i$ of \eqref{eq:condstates}; then, the no-broadcasting Theorem~\ref{thm:nobroad} will allow us to conclude that all the $\rho^B_i$ can be diagonalized in a same basis, which is equivalent to saying that $\rho_{AB}$ is classical on $B$.

That $\Gamma_{B\rightarrow B_1B_2}$ broadcasts the individual states $\rho^B_i$ is true, since we have (we focus on the $B_1$ output for concreteness, but the same goes for $B_2$):
\[
\begin{aligned}
\Tr_{B_2}(\Gamma_{B\rightarrow B_1B_2}[\rho^B_i])
&= \frac{1}{p_i}\Tr_{AB_2}((E^A_i\otimes\I_{B_1B_2}) (\idmap_A\ot \Gamma_{B\rightarrow B_1B_2})[\rho_{AB}])\\
&= \frac{1}{p_i}\Tr_{A}((E^A_i\otimes\I_{B_1})\tilde{\rho}_{AB_1})\\
&= \frac{1}{p_i}\Tr_{A}((E^A_i\otimes\I_{B})\rho_{AB})\\
&= \rho^B_i.
\end{aligned}
\]
The first equality is due to the definition of conditional state \eqref{eq:condstates}, and the fact that the POVM $\{E_i^A\}$ and the map $\Gamma_{B\rightarrow B_1B_2}$ operate on different systems; the second and third equalities are due to the broadcasting conditions. \qed
\end{proof}

\subsection{No-local-broadcasting}

We are now in the position to give a straightforward proof of Theorem~\ref{thm:nolocal}.

\begin{proof} (of Theorem~\ref{thm:nolocal}) That a classical-classical state can be locally broadcast is trivial. To prove the other direction, let us assume that $\rho_{AB}$ can be locally broadcast, and let $\Lambda_{A\rightarrow A_1A_2}$ and $\Gamma_{B\rightarrow B_1B_2}$ be the locally broadcasting maps, so that $\tilde{\rho}_{A_1A_2B_2B_2}=(\Lambda_{A\rightarrow A_1A_2}\otimes \Gamma_{B\rightarrow B_1B_2})[\rho_{AB}]$ satisfies the broadcasting conditions \eqref{eq:localbroadcond}. We will prove that $\rho_{AB}$ can also be locally broadcast on both $A$ and $B$, and hence it is classical on both $A$ and $B$, which means it is classical-classical. For the sake of concreteness we will focus on proving that $\rho_{AB}$ can be locally broadcast on $B$. A similar proof can be followed to prove classicality on $A$.

Besides $\rho_{AB}$ and $\tilde{\rho}_{A_1A_2B_1B_2}$ defined above, it is convenient to consider also
\[
\tilde{\rho}'_{AB_1B_2}:=(\idmap_A\otimes \Gamma_{B\rightarrow B_1B_2})[\rho_{AB}].
\]
Notice that it holds
\[
\tilde{\rho}_{A_1A_2B_1B_2}=(\Lambda_{A\rightarrow A_1A_2}\otimes \idmap_B)[\tilde{\rho}'_{AB_1B_2}].
\]
The  key point is that one goes from $\rho_{AB}$, to $\tilde{\rho}'_{AB_1B_2}$, to $\tilde{\rho}_{A_1A_2B_1B_2}$ by a sequence of local operations. This, together to monotonicity of mutual information  under local operations (including partial trace), implies
\[
I(A_1:B_1)_{\tilde{\rho}} \leq I(A_1A_2:B_1)_{\tilde{\rho}} \leq I(A:B_1)_{\tilde{\rho}'} \leq I(A:B_1B_2)_{\tilde{\rho}'} \leq I(A:B)_\rho.
\]
Notice that, because of the broadcasting conditions, the leftmost quantity is actually equal to the rightmost quantity, hence, all the mutual information quantities in the latter equation are equal. In particular, $I(A:B_1)_{\tilde{\rho}'}=I(A:B)_\rho$. Moreover,
\[
\tilde{\rho}'_{AB_1}=\big(\idmap_{A}\otimes\Gamma'_{B\rightarrow B_1}\big)[\rho_{AB}],
\]
with $\Gamma'_{B\rightarrow B_1}=(\idmap_{B_1}\otimes\Tr_{B_2})\circ \Gamma_{B\rightarrow B_1B_2}$. It follows then from Theorem \ref{thm:originalfawzirenner} that there is a recovery channels $R^{(1)}_{B_1\rightarrow B}$ such that
\[
(\idmap_A\otimes (R^{(1)}_{B_1\rightarrow B} \circ \Gamma'_{B\rightarrow B_1}))[\rho_{AB}] =\rho_{AB}
\]
One can argue similarly about $B_2$. We  arrive at the conclusion that there are two channels $R^{(1)}_{B_1\rightarrow B_1}$ and $R^{(2)}_{B_2\rightarrow B_2}$ such that
\[
(R^{(1)}_{B_1\rightarrow B_1}\otimes R^{(2)}_{B_2\rightarrow B_2})\circ \Gamma_{B\rightarrow B_1B_2}
\]
locally broadcasts $\rho_{AB}$ on $B$. \qed
\end{proof}

\section{Brodcasting mutual information}

The characterization of recoverability in terms of mutual information captured by Theorem \ref{thm:originalfawzirenner} is such---one could  say, `strong enough'---that the local (or unilocal) broadcasting of quantum states, understood in a structural sense---that is, in terms the density matrices---is fully equivalent to the broadcasting of the correlations contained in the states, as quantified by mutual information. More precisely, one can state a no-go theorem for local broadcasting, which, at least at face value, is more general than Theorem \ref{thm:nolocal}.
\begin{theorem}
Let $\rho_{AB}$ be a bipartite state. There exist local maps $\Lambda_{A\rightarrow A_1A_2}$ and $\Gamma_{B\rightarrow B_1B_2}$ such that
\beq
\label{eq:broadcondmutual}
\tilde{\rho}_{A_1A_2B_2B_2}=(\Lambda_{A\rightarrow A_1A_2}\otimes\Gamma_{B\rightarrow B_1B_2})[\rho_{AB}]
\eeq
satisfies
\beq
I(A_1:B_1)_{\tilde{\rho}}=I(A_2:B_2)_{\tilde{\rho}}=I(A:B)_{\rho_{AB}}
\eeq
if and only if $\rho_{AB}$ is classical-classical.
\end{theorem}
This version of the no-local-brodcasting theorem would indeed appears to be more general than Theorem \ref{thm:nolocal} because if  $\tilde{\rho}_{A_1A_2B_2B_2}$ satisfies the broadcasting conditions \eqref{eq:localbroadcond}, then it satisfies also conditions \eqref{eq:broadcondmutual}, but the opposite is not immediately evident. Nonetheless, it is true---and we know it thanks to Theorem \ref{thm:originalfawzirenner}. Similarly, one can have an alternative version of the no-unilocal-brodcasting theorem.
\begin{theorem}
\label{thm:unilocalmutual}
Let $\rho_{AB}$ be a bipartite state. There exists a local map $\Gamma_{B\rightarrow B_1B_2}$  such that
\beq
\tilde{\rho}_{AB_1B_2}=(\idmap_A\otimes\Gamma_{B\rightarrow B_1B_2})[\rho_{AB}]
\eeq
satisfies
\beq
I(A:B_1)_{\tilde{\rho}}=I(A:B_2)_{\tilde{\rho}}=I(A:B)_{\rho}.
\eeq
if and only if $\rho_{AB}$ is classical on $B$.
\end{theorem}

\section{Quantifying non-classical correlations through broadcasting}

So far we have characterized quantum correlations only qualitatively, via no-go theorems. Moreover, we have made use of Theorem~\ref{thm:originalfawzirenner} only for the case of exact recoverability. In this section, we delineate some ways in which local and unilocal broadcasting can be used to \emph{quantify} the degree of non-classicality of correlations of the state $\rho_{AB}$ under scrutiny. Interestingly, we will connect broadcasting to the well established quantifier of non-classical correlations known as quantum discord. The latter is an asymmetric quantity, based on the notion of minimal loss of correlations, as measured by mutual information, when one tries to `extract' such correlations and map them into a classical register. More explicitly, consider quantum-to-classical channels $\mathcal{M}_{B\rightarrow B'}[\sigma_B]=\sum_i \Tr(M_i\sigma_B)\proj{i}_{B'}$, where $\{M_i\}$ is a POVM on $B$, and $\{\ket{i}\}$ is an orthonormal basis for the `classical register' $B'$.
We define $\rho'_{AB'} = (\idmap_A\ot \mathcal{M}_{B\rightarrow B'})[\rho_{AB}]$. Then, the discord of $\rho_{AB}$ on $B$ is equal to 
\beq
\label{eq:discord}
D(A|B)_\rho := I(A:B)_\rho - \max_{\mathcal{M}_{B\rightarrow B'}}I(A:B')_{\rho'},
\eeq
Here the maximum is over all quantum-to-classical channels. Notice that discord can be rewritten as
\beq
D(A|B)_\rho := \min_{V_{B\rightarrow B'E}} I(A:E|B')_{V\rho V^\dagger},
\eeq
where $V = V_{B\rightarrow B'E}$ is any isometry that realizes a quantum-to-classical channel  $\mathcal{M}_{B\rightarrow B'}$. Discord vanishes only for the quantum-classical states~\cite{ollivier2001quantum,henderson2001classical,hayashi2006quantum}; in all other cases, there is a loss of correlations---a measured by quantum mutual information---in the local quantum-to-classical mapping. Nonetheless, one can try to recover $\rho_{AB}$ from the quantum-classical state $\rho_{AB'}$, via a recovery channel $R_{B'\rightarrow B}$. It is easy to check that, without loss of generality, such a recovery map consists of a preparation procedure, so that the composition of measurement $\mathcal{M}$ and preparation/recovery $R$ gives rise to a so-called entanglement breaking map, $R_{B'\rightarrow B}\circ\mathcal{M}_{B\rightarrow B'}[\sigma_B]=\Lambda^\textrm{EB}_{B}[\sigma_B]=\sum_i \Tr(M_i\sigma_B) \tau_B^i$, for all $\sigma_B$, with $\{\tau_B^i\}$ a collection of states. Let us define~\cite{seshadreesan2014fidelity}
\beq
F^{\textrm{EB}}_{B}(\rho_{AB}) := \max_{\Lambda^\textrm{EB}_{B}}F\big(\rho_{AB},(\idmap_A\otimes\Lambda^\textrm{EB}_{B})[\rho_{AB}]\big).
\eeq
Then, Theorem~\ref{thm:originalfawzirenner} implies~\cite{seshadreesan2014fidelity}
\beq
\label{eq:wildebound}
D(A|B)_\rho \geq -2 \log_2 F^{\textrm{EB}}_{B}(\rho_{AB}).
\eeq

\subsection{Imperfect structural local broadcasting}

Although cloning and broadcasting of general unknown states is not possible, as formalized by the no-cloning and no-broadcasting theorems, one can consider how well the task can be achieved, at least in an approximate sense. This corresponds to the relative large topic of optimal (albeit not perfect) quantum cloners (see, e.g.,~\cite{gisin1997optimal,wernercloning}). The same applies to local or unilocal broadcasting. For example, one can consider the state-dependent~\footnote{A large part of the study about optimal cloners deals with single system, where it is natural to discuss the cloning of a set of states, or universal cloners, where the figure of merit is either an average, or state-independent (for pure states).} maximally achievable fidelity
\beq
\label{eq:approxbroad}
F^{\textrm{max}}_{B,1\rightarrow 2}(\rho_{AB}):=\max_{\Lambda_{B\rightarrow B_1B_2}} F\big(\rho_{AB},\Tr_{B_1}((\idmap_A\otimes\Lambda_{B\rightarrow B_1B_2})[\rho_{AB}])\big),
\eeq
where the maximum is over maps $\Lambda_{B\rightarrow B_1B_2}$ whose output is invariant under swap of $B_1B_2$.
Alternatively, one could consider the average of the fidelities, for a general map that does not have symmetric output, but one can argue that a symmetric output is always optimal, thanks to the (joint) concavity of the fidelity in each of its arguments~\cite{nielsen2010quantum}.
Notice that, exactly because of the symmetry of the output of $\Lambda_{B\rightarrow B_1B_2}$, on the right side of  \eqref{eq:approxbroad} we can indifferently consider the trace over $B_1$ or $B_2$. The no-unilocal-brodcasting Theorem~\ref{thm:unilocalbroad} ensures that $F^{\textrm{max}}_{B,1\rightarrow 2}(\rho_{AB})<1$ (strictty) as soon as $\rho_{AB}$ is not classical on $B$.

We now observe that any entanglement breaking map can be seen as the composition of a map with symmetric output followed by a partial trace, because
\[
\sum_i \Tr(M_i\sigma_B) \tau_B^i = \Tr_{B_1}\left(\sum_i \Tr(M_i\sigma_B) \tau_{B_1}^i\otimes\tau_{B_2}^i\right).
\]
This implies that $F^{\textrm{max}}_{B,1\rightarrow 2}(\rho_{AB})\geq F^{\textrm{EB}}_{B}(\rho_{AB})$, which combined with \eqref{eq:wildebound}, gives
\[
D(A|B)_\rho \geq -2 \log_2 F^{\textrm{max}}_{B,1\rightarrow 2}(\rho_{AB}).
\]
Thus, we see that discord can be bounded in terms of the quality of approximate broadcasting. In Ref.~\cite{piani2015hierarchy} this relation is further explored, showing, on one hand, that considering a larger number of outputs leads to more stringent bound on the discord, and, on the other hand, that each quantities like $F^{\textrm{max}}_{B,1\rightarrow 2}(\rho_{AB})$ can be computed numerically in an efficient and reliable way, since such a quantity can be calculated by semidefinite programming~\cite{boyd2009convex,watrous2012simpler}.

\subsection{Imperfect local broadcasting of mutual information}

In the same way in which we have cast the no-local-broadcasting and no-unilocal-brodcasting theorem in terms of broadcasting mutual information, so we can approach the issue of approximate broadcasting. We will focus on unilocal broadcasting. Then, for any channel $\Lambda_{B\rightarrow B'^n}$, where $B'^n = B'_1B'_2\ldots B'_n$, and each $B'_1$, $B'_2$, ..., $B'_n$ could potentially have a dimensionality different from that of $B$, one can define the average loss of correlations in broadcasting as
\beq
I(A:B)_{\rho_{AB}} - \frac{1}{n}\sum_{i=1}^n I(A:B'_i)_{(\idmap_A\ot \Lambda_{B\rightarrow B'^n})[\rho_{AB}]}
\eeq
The mutual information version of the no-unilocal-broadcasting theorem, Theorem~\ref{thm:unilocalmutual}, says that such an average loss is strictly positive for all $n\geq2$, if $\rho_{AB}$ is not quantum-classical. One proves that such an average loss actually converges to exactly the quantum discord \eqref{eq:discord}, with $n$ going to infinity~\cite{brandao2013quantum}.

\section{Conclusions}

It is hard to overestimate the importance of the no-cloning theorem in our understanding of quantum mechanics and quantum information, and the pivotal role it has played in the latter field. For example, born from the attempts to reconcile entanglement with a principle of no-faster than light signalling, it contributed to the development of quantum cryptography. We have seen that the counterpart of no-cloning in the scenario where one considers mixed states is no-broadcasting.

When it comes to distributed system, and to the study of the limitations in the local manipulation of correlations, other no-go theorems can be derived, like the no-local-broadcasting theorem or the no-unilocal-broadcasting theorem. It is worth emphasizing one last time that the latter no-go results apply to single distributed quantum states, contrary to the no-cloning and no-broadcasting theorems, which instead deal with multiple states. 

The quantification of the limits in the local manipulation of correlations provide  a sound and physically meaningful way to quantify the non-classicality of correlations. Interestingly, this field of research connects directly to recent and exciting advances in our understanding of quantum recoverability. The latter connection can also be exploited for an efficient numerical approach to the quantification of the quantumness of correlations, allowing, for example, the derivation of reliable numerical bounds to quantum discord.

\section*{Acknowledgements} The productions of these notes was supported by the European Union's Horizon 2020 research and innovation programme under the Marie Sklodowska-Curie grant agreement No 661338. I would like to thank the Institute for Quantum Computing at the University of Waterloo for its hospitality during the completion of these notes.


\end{document}